\documentclass[12pt,preprint]{aastex}

\slugcomment{}

\shorttitle{{\it Swift} AGN}
\shortauthors{Ajello et al.}

\usepackage{calrsfs,euscript,mathrsfs}

\begin{document}


\title{The 60-month all-sky BAT Survey of AGN and the Anisotropy
of Nearby AGN}


\author{
M.~Ajello\altaffilmark{1},  D.~M.~Alexander\altaffilmark{2}, J.~Greiner\altaffilmark{3},
G.~M.~Madejski\altaffilmark{1}, N.~Gehrels\altaffilmark{4} and D.~Burlon\altaffilmark{3} 
}
\altaffiltext{1}{Kavli Institute for Particle Astrophysics and Cosmology, Department of Physics and SLAC National Accelerator Laboratory, Stanford University, Stanford, CA 94305, USA}
\altaffiltext{2}{Department of Physics, Durham University, Durham DH1 3LE, UK}
\altaffiltext{3}{Max Planck Institut f\"{u}r Extraterrestrische Physik, P.O. Box 1603, 85740, Garching, Germany}
\altaffiltext{4}{NASA Goddard Space Flight Center, Greenbelt, MD 20771, USA}

\begin{abstract}
Surveys above 10\,keV represent one of the 
the best resources to provide an unbiased census
of the population of Active Galactic Nuclei (AGN).
We present the results of 60\,months of observation of the hard X-ray
sky with {\it Swift}/BAT. In this timeframe, BAT detected (in the 15--55\,keV band)
720 sources in an all-sky survey of which 428 are associated with AGN,
most of which are nearby. Our sample has negligible incompleteness
and statistics a factor of $\sim$2 larger over similarly complete sets of AGN. Our sample contains (at least) 15 bona-fide Compton-thick AGN and 3 likely
candidates. Compton-thick AGN represent a $\sim$5\,\% of AGN samples
detected above 15\,keV.
We use the BAT dataset to refine the determination of the LogN--LogS of AGN which
is extremely important, now that NuSTAR prepares for launch,
towards assessing the AGN contribution to the cosmic X-ray
background. We show that the LogN--LogS of AGN selected above 10\,keV is 
now established to a $\sim$10\,\% precision.
We derive the luminosity function of Compton-thick AGN and measure a space density  of 7.9$^{+4.1}_{-2.9}\times10^{-5}$\,Mpc$^{-3}$
for objects with a de-absorbed luminosity
larger than 2$\times10^{42}$\,erg s$^{-1}$.
As the BAT
AGN are all mostly local, they allow us to investigate the spatial distribution
of AGN in the nearby Universe regardless of absorption.
We find concentrations of AGN that coincide spatially with the largest
congregations of matter in the local ($\leq$85\,Mpc) Universe.
There is some evidence that the fraction of Seyfert 2 objects is larger
than average in the direction of these dense regions.
\end{abstract}


%
%
%
\section{Introduction}

There is a general consensus that the cosmic X-ray
background (CXB), discovered more than 40 years ago \citep{giacconi62},
is produced by integrated emission of 
Active Galactic Nuclei (AGN). 
Indeed, below $\sim$3\,keV sensitive observations with Chandra and XMM-Newton
have directly resolved as much as 80\,\% of the CXB into AGN \citep{worsley05,luo11}. However, above  5\,keV, due to the lack of sensitive observations,
most of the CXB emission is at present unresolved.
Population synthesis models have successfully shown, 
in the context of the AGN unified theory \citep{antonucci93},
that  AGN  with various level of obscuration and at different redshifts 
can account for  80--100\% of the CXB up to $\sim$100\,keV
\citep{comastri95,gilli01,treister05}.
In order to reproduce the spectral shape and the intensity of the CXB,
these models require that Compton-thick AGN 
(N$_{H}\geq1.4\times10^{24}$\,cm$^{-2}$) contribute  $\sim$10\,\% of the total
CXB intensity.
With such heavy absorption Compton-thick AGN have necessarily to be numerous,
comprising perhaps up to 30--50\,\% of the AGN population in the local Universe \citep[e.g.][]{risaliti99}. However, it is still surprising that only a very 
small fraction of the population of Compton-thick AGN has been uncovered
so far \citep[][and references therein]{comastri04,dellaceca08a}.

Studies of AGN are best done above 10\,keV where the nuclear radiation
pierces through the torus for all but the largest column densities.
Focusing optics like those mounted on NuSTAR and ASTRO-H \citep[respectively,][]{harrison10,takahashi10} will allow us to reach, for the first time, sensitivities $\leq$10$^{-13}$\,erg cm$^{-2}$ s$^{-1}$ above 10\,keV permitting us 
 to resolve a substantial fraction of the CXB emission
in this band. Given their small field of views (FOVs) those instruments
will need large exposures in order to gather reasonably large AGN samples.
Because of their good sensitivity the AGN detected
by NuSTAR and ASTRO-H should be at redshift $\sim$1, 
but, due to the small area surveyed,
very few if any will be  at much lower redshift.

All-sky surveys, like those performed by {\it Swift}/BAT and INTEGRAL above
10\,keV are very effective in making a census of nearby AGN, 
thus providing a natural extension to more sensitive 
(but with a narrower FOV) missions. Here we report on the all-sky
sample of AGN detected by BAT in 60\,months of exposure. Our sample comprises
428 AGN detected in the whole sky and represents  
a factor of $\sim$2 improvement in number statistics when compared to 
previous complete samples
\citep[e.g.][]{burlon11}. In this paper, we present the sample
and refine the determination of the source count distribution and of the luminosity
function of AGN. This is especially important considering the upcoming launch of 
NuSTAR (scheduled for March~2012) as it allows us to make accurate predictions
for the expected space densities of distant AGN. We also use the BAT sample
to investigate the spatial distribution of AGN in the local Universe.
We leave for an upcoming publication the follow-up of  all new sources using 
2--10\,keV data and the determination of the absorption distribution.

This paper is organized as follows:  the BAT observations are discussed
in $\S$~\ref{sec:obs}, while $\S$~\ref{sec:counts} and $\S$~\ref{sec:xlf}
discuss respectively the source count distribution and the luminosity function
of AGN. In $\S$~\ref{sec:aniso} we present a measurement of the over-density
of AGN in the local Universe, while in $\S$~\ref{sec:nustar} the prospects
for the detection of AGN by NuSTAR are discussed in the framework of 
the BAT observations and population synthesis models.
Finally, $\S$~\ref{sec:concl} summarizes our findings.
Throughout this paper, we assume a standard concordance cosmology 
(H$_0$=71\,km s$^{-1}$ Mpc$^{-1}$, $\Omega_M$=1-$\Omega_{\Lambda}$=0.27).

%
%
%
\section{Properties of the Sample}
\label{sec:obs}
The Burst Alert Telescope \citep[BAT;][]{barthelmy05}
onboard the {\em Swift} satellite \citep{gehrels04}, represents 
a major improvement in sensitivity for imaging of the hard X-ray sky. 
BAT is a coded mask telescope with a wide field of view 
(FOV, 120$^\circ\times$90$^{\circ}$ partially coded) aperture sensitive in
the 15--200\,keV range. Thanks 
to its wide FOV and its pointing strategy, BAT monitors continuously
up to 80\% of the sky every day achieving, after several years, deep exposures across the entire sky.
Results of the BAT survey \citep{markwardt05,ajello08a,tueller08}
show that BAT reaches a sensitivity of $\sim$1\,mCrab\footnote{1\,mCrab in the 15--55\, keV band corresponds to 
1.27$\times10^{-11}$\,erg cm$^{-2}$ s$^{-1}$}
in 1\,Ms of exposure. 
Given its sensitivity and the large exposure already accumulated in 
the whole sky, BAT is  an excellent instrument for studying
populations whose emission is faint in  hard X-rays.

For the analysis presented here we use 60\,months of 
 {\em Swift}/BAT observations taken 
between March 2005 and March 2010. Data screening and
processing was performed according to the recipes presented in \cite{ajello08a} and
\cite{ajello08b}.
The chosen energy interval is 15--55\,keV. The all-sky image is obtained as  
the weighted average of all the shorter observations.
The final image shows a Gaussian normal noise and we identified source candidates
as those excesses with a signal-to-noise (S/N) ratio $\geq$  5\,$\sigma$. The final sample comprises 720 sources detected all-sky.
Identification of these objects was performed by cross-correlating our
catalog with the catalogs of \cite{tueller08}, \cite{cusumano10}, \cite{voss10}, and
 \cite{burlon11}. Whenever available we used the newest optical
identifications provided by \cite{masetti08}, \cite{masetti09}, and \cite{masetti10}.
Of the 720 all-sky sources only 37 (i.e. $\sim$5\,\%) do not have a firm identification.
This small incompleteness does not change when excluding or including
the Galactic plane.  
Of the 720 objects, 428 are identified with AGN. This represents
an improvement of a factor $>$2 in the number of detected AGN 
with respect to previous complete samples \citep[e.g.][]{ajello09b,burlon11}.
\cite{cusumano10} recently reported on the sample of sources detected by BAT 
in 58\,months of observations. Their catalog is constructed using three
energy bands and selecting $\geq$4.8\,$\sigma$ excesses in any of the three bands.
As such their catalog is larger than the one presented here. However, for the 
scope of this and future analyses \citep[e.g. a follow-up work 
of that presented in ][]{burlon11} 
it is important to have a clean sample whose selection effects are well understood
and can be accounted for during the analysis.

\begin{figure}[ht!]
  \begin{center}
  	 \includegraphics[scale=0.60]{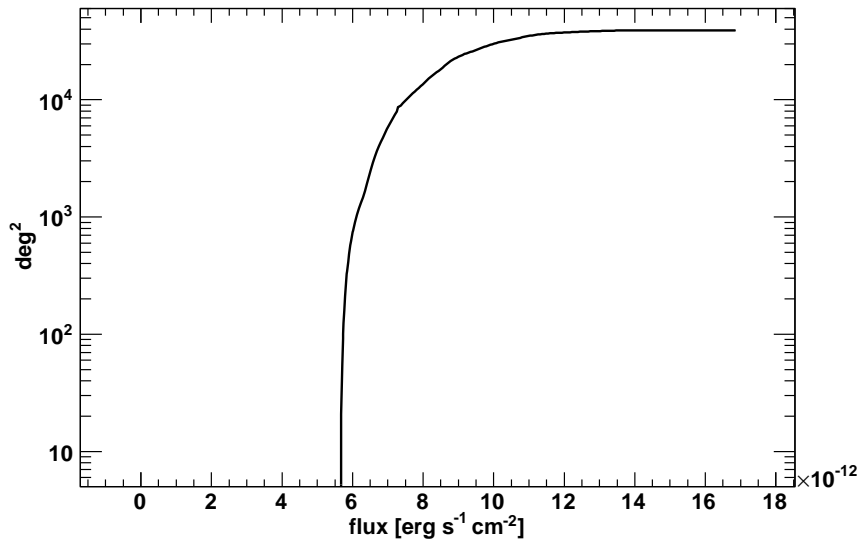}
  \end{center}
  \caption{Sky coverage of the BAT survey for the 15--55\,keV band and
for sources detected all-sky above the 5\,$\sigma$ level.
\label{fig:skycov}}
\end{figure}

Fig.~\ref{fig:skycov} shows the sky coverage of the BAT
survey. It is apparent that the limiting flux is $\sim$0.45\,mCrab ($\sim$5.5$\times10^{-12}$\,erg cm$^{-2}$ s$^{-1}$) and that the BAT survey becomes complete (for the whole sky)
for source fluxes $\geq$1\,mCrab. The sensitivity  scales nicely with the inverse
of the  square root of the exposure time as testified by the limiting sensitivity
of 0.6\,mCrab reached in 36\,months of observations \citep{ajello09}.

\begin{deluxetable}{lcccclccccc}
\tablewidth{0pt}
\tabletypesize{\scriptsize}
\rotate
\tablecaption{The 428 AGN detected by BAT \label{tab:cat}\tablenotemark{a}.}
\tablehead{
\colhead{SWIFT NAME}       & \colhead{R.A.}                &
\colhead{Decl.}            & \colhead{Position Error}      &
\colhead{Flux}                &
\colhead{S/N}              & \colhead{ID}                  &
\colhead{Type\tablenotemark{b}}             & \colhead{Redshift}      &
\colhead{Photon Index}     & \colhead{Log L$_{X}$}\\
%
%
\colhead{}                        & \colhead{\scriptsize (J2000)}         &
\colhead{\scriptsize (J2000)}     & \colhead{\scriptsize (arcmin)}     & 
\colhead{\scriptsize(10$^{-11}$ cgs)} &
\colhead{}                        & \colhead{}                            &
\colhead{}                        & \colhead{} &
\colhead{}                        & \colhead{} 
}
\startdata
J0004.2+7018 & 1.050 & 70.300 & 6.551 & 0.76 & 5.5 & 2MASX J00040192+7019185 & AGN  & 0.0960 & 2.04$\pm0.46$ & 44.2\\ 
J0006.2+2010 & 1.571 & 20.168 & 3.276 & 1.06 & 6.5 & Mrk 335 & Sy1  & 0.0254 & 2.60$\pm0.32$ & 43.2\\ 
J0010.4+1056 & 2.622 & 10.947 & 2.281 & 1.85 & 11.0 & QSO B0007+107 & BLAZAR  & 0.0893 & 2.23$\pm0.20$ & 44.6\\ 
J0018.9+8135 & 4.732 & 81.592 & 4.720 & 0.94 & 6.5 & QSO J0017+8135 & BLAZAR  & 3.3600 & 2.51$\pm0.52$ & 48.3\\ 
J0021.2-1908 & 5.300 & -19.150 & 4.773 & 0.92 & 5.1 & 1RXSJ002108.1-190950 & AGN  & 0.0950 & 1.96$\pm0.45$ & 44.3\\ 
J0025.0+6826 & 6.264 & 68.436 & 4.721 & 0.78 & 5.6 & IGR J00256+6821 & Sy2  & 0.0120 & 1.66$\pm0.33$ & 42.4\\ 
J0033.4+6125 & 8.351 & 61.431 & 4.274 & 1.01 & 7.3 & IGR J00335+6126 & AGN  & 0.1050 & 2.46$\pm0.28$ & 44.5\\ 
J0034.6-0423 & 8.651 & -4.400 & 6.165 & 0.92 & 5.2 & 2MASX J00343284-0424117 & AGN  & 0.0000 & 1.79$\pm0.43$ & \nodata\\ 
J0035.8+5951 & 8.965 & 59.852 & 2.095 & 2.05 & 14.8 & 1ES 0033+59.5 & BLAZAR  & 0.0860 & 2.74$\pm0.18$ & 44.6\\ 
J0038.5+2336 & 9.648 & 23.600 & 5.132 & 1.01 & 6.2 & Mrk 344 & AGN  & 0.0240 & 1.80$\pm0.59$ & 43.1\\ 
J0042.8-2332 & 10.701 & -23.548 & 3.068 & 2.52 & 14.7 & NGC 235A & Sy2  & 0.0222 & 1.90$\pm0.11$ & 43.4\\

\enddata
\tablenotetext{a}{The full table is available in the online version of the
paper.}
\tablenotetext{b}{AGN are sources lacking an exact optical classification.}
\end{deluxetable}

\begin{figure}[ht!]
  \begin{center}
  	 \includegraphics[scale=0.8]{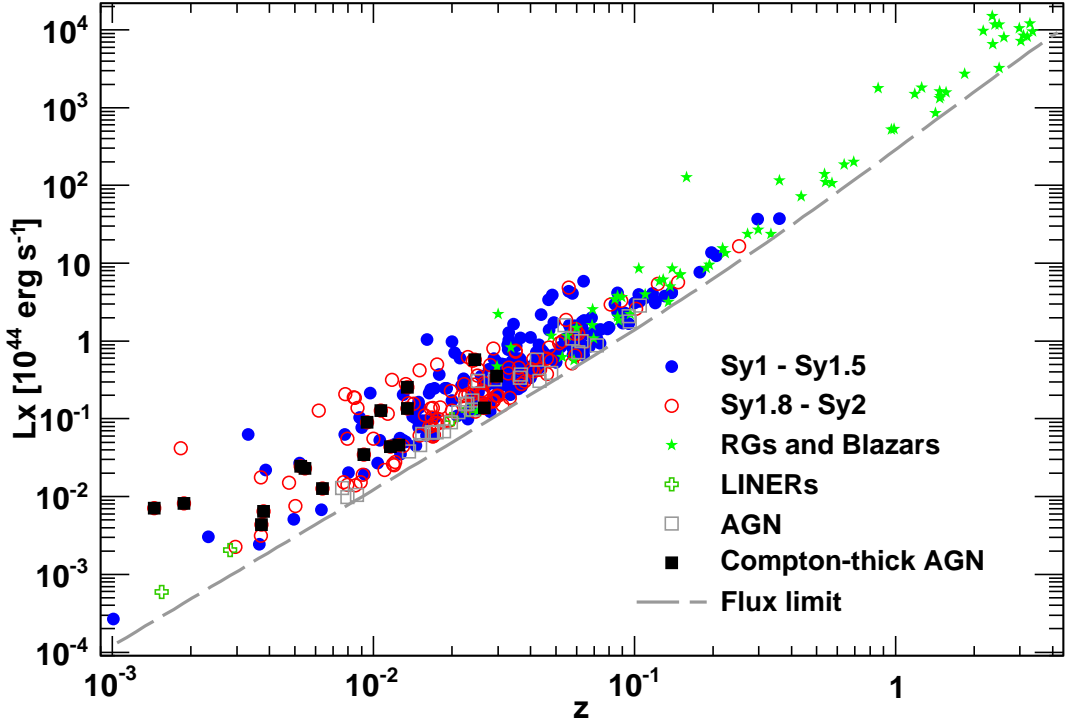}
  \end{center}
  \caption{Position on the luminosity-redshift plane of the 428 AGN
detected by BAT in the 15--55\,keV band. The color coding reflects the optical classification reported
in Tab.~\ref{tab:cat}. AGN are sources lacking an exact optical classification.
The black squares mark the position of the Compton-thick AGN reported
in Tab.~\ref{tab:ct}. Note that their luminosities were not corrected
for absorption (see text for details). The dashed line shows the flux limit
of the BAT survey of 5.5$\times10^{-12}$\,erg cm$^{-2}$ s$^{-1}$.
\label{fig:sample}}
\end{figure}

%
%
\subsection{Jet-dominated and Disk-dominated Objects}
Jet-dominated AGN (radio galaxies and blazars) constitute a  $\sim$15\,\%
fraction of the BAT samples \cite[e.g. see][]{ajello09b}. This is confirmed
also here where 67 (out of the 428 AGN) are classified as either radio-galaxies
or blazars.
The remaining 361 AGN are associated with objects optically
classified as Seyfert galaxies (323 objects) or with nearby
galaxies (38 sources), through the detection of a soft X-ray counterpart, for which 
an optical classification is not yet available.
The full sample is reported in Tab.~\ref{tab:cat}.
For all the sources, $k-$corrected $L_X$ luminosities  were computed 
according to:
\begin{equation}
L_X\, =\, 4\pi d_{\rm L}^2 {F_X \over (1+z)^{2-\Gamma_X} } 
\label{lx}
\end{equation}
where $F_X$ is the X-ray energy flux in the 15--55\,keV band and
$\Gamma_X$ is the photon index. This assumes that the source spectra  
are adequately well described by a power law in the 15--55\,keV 
band in agreement with what found by \cite{ajello08b}, \cite{tueller08}, and
\cite{burlon11}. Unless noted otherwise (i.e. $\S$~\ref{sec:cthick_dens})
luminosities are not corrected for absorption along the line
of sight since this correction is different than unity (in the 15--55\,keV band)
only for Compton-thick AGN \citep[see Fig.~11 in][]{burlon11} and
does not introduce any apparent bias in any of the results shown
in the next sections.

Fig.~\ref{fig:sample} shows the position of the 428 sources in the luminosity-redshift
plane for the different optical classifications reported in Tab.~\ref{tab:cat}.
The BAT AGN sample spans almost 8 decades in luminosity and includes sources detected
from z$\approx$0.001 (i.e. $\sim$4\,Mpc) up to z$\approx$4.
It is also evident that Seyfert galaxies
dominate the low-luminosity part of the sample, while  blazars and radio-galaxies
dominate the high-luminosity part of the sample.
The increased exposure of BAT allows us to detect fainter AGN with
respect to previous samples. Indeed,
the average flux of the Seyfert-like AGN decreased by $\sim$20\,\%
when comparing it to the sample of AGN reported in \cite{burlon11}.
 Since the average redshift in the two samples is very similar,
this translates into a larger number of low-luminosity AGN.

%
%
\subsection{Compton-thick AGN}
Hard X-ray selected samples are among the best resources to uncover
Compton-thick AGN which are otherwise difficult to detect.
A detailed measurement of the absorbing column density of 
all the AGN in this sample is beyond the scope of this paper and left
for a future publication.
However,  in order to determine the likely candidates,
it is possible to cross-correlate our source list with catalogs
of Compton-thick AGN. Our AGN catalog contains all the 9 Compton-thick
AGN reported by \cite{burlon11} and 6 additional Compton-thick AGN
reported in the list of bona-fide objects of \cite{dellaceca08a}.
There 3 additional sources which are labeled as Compton-thick candidates
by \cite{dellaceca08a} (see their Table~2) which are also detected in this sample.
The full list of 18 known Compton-thick AGN contained in this sample is reported
in Tab.~\ref{tab:ct}. It is clear that the number of (likely) Compton-thick
AGN is doubled with respect to the sample of \cite{burlon11} and
that Compton-thick AGN represent a `steady' 5\,\%  fraction (i.e. $\sim$18/361)
of AGN samples selected above 10\,keV. 

The redshift distribution of Compton-thick AGN is also different than
that of the whole AGN sample.
The median redshift of the Compton-thick AGN 
of Tab.~\ref{tab:ct} is 0.010 while that one of the entire AGN
sample is 0.029. Compton-thick AGN can be detected by BAT only within
a distance of $\sim$100\,Mpc beyond which the strong flux suppression
caused by the Compton-thick medium
limits the capability of BAT to detect these objects.

We also checked if any of the remaining
3  bona-fide Compton-thick AGN (or the 20 remaining 
candidates) reported in \cite{dellaceca08a}
lie just below the reliable BAT detection threshold.
 None of the remaining
sources in the above lists exhibits a significance larger than
3.5\,$\sigma$ in our analysis. 
This means that none of these sources  are likely to be
detectable
by BAT in a deeper survey. The main consequence is however that the
new Compton-thick objects that will appear in the BAT samples will
be new (i.e. previously un-studied) sources. A few might already
be present in this sample and this aspect will be investigated in
a follow-up study.

\begin{deluxetable}{llcccc}
\tablewidth{0pt}
\footnotesize
\tablecaption{Known Compton-thick AGN detected in the BAT sample.
Unless written explicitly the values of the absorbing column density
come from \cite{burlon11}.
\label{tab:ct}}
\tablehead{\colhead{NAME} & \colhead{Type} & \colhead{Reshift} &
\colhead{R.A.} & \colhead{Decl.} & \colhead{N$_H$}\\
\colhead{}         & \colhead{}   & \colhead{}     
& \colhead{\scriptsize (J2000)}    & \colhead{\scriptsize (J2000)} &
\colhead{($10^{24}$ cm$^{-2}$)}
}
\startdata
 
  NGC 424            &Sy2 & 0.011588 &  17.8799 &  -38.0944 & 1.99 \\
  NGC 1068           &Sy2 & 0.003787 &  40.7580 &   -0.0095 & $>$10\\
	 NGC 1365\tablenotemark{a,b}           &Sy1.8&0.005460 &  53.4442 &  -36.1292 & 3.98\\
  CGCG 420-015       &Sy2 & 0.029621 &  73.3804 &    4.0600 & 1.46 \\
  SWIFT J0601.9-8636 &Sy2 & 0.006384 &  91.1972 &  -86.6245 & 1.01\\
  Mrk 3              &Sy2 & 0.013509 &  93.9722 &   71.0311 & 1.27\tablenotemark{e}\\
  UGC 4203\tablenotemark{a,c}           &Sy2 & 0.013501 & 121.0585 &    5.1217 & $>$1.00\tablenotemark{e}\\
  NGC 3079           &Sy2 & 0.003720 & 150.4701 &   55.6978 & 5.40\\
  NGC 3281           &Sy2 & 0.010674 & 157.9743 &  -34.8571 & 1.96\tablenotemark{e}\\
  NGC 3393           &Sy2 & 0.012500 & 162.1000 &  -25.1539 & 4.50\\
  NGC 4939           &Sy1 & 0.010374 & 196.1000 &  -10.3000 & $>$10\tablenotemark{e}\\
  NGC 4945           &Sy2 & 0.001878 & 196.3726 &  -49.4742 & 2.20\tablenotemark{e}\\
  Circinus Galaxy    &Sy2 & 0.001447 & 213.3828 &  -65.3389 & 4.30\tablenotemark{e}\\
  NGC 5728           &Sy2 & 0.009467 & 220.6916 &  -17.2326 & 1.0\\
  ESO 138-1          &Sy2 & 0.009182 & 253.0085 &  -59.2386 & 1.5\tablenotemark{e,f}\\
  NGC 6240           &Sy2 & 0.024480 & 253.3481 &    2.3999 & 1.83\\
  NGC 6552\tablenotemark{a,d}           &Sy2 & 0.026550 & 270.0981 &   66.6000 & $>$1.00\tablenotemark{e} \\
  NGC 7582           &Sy2 & 0.005253 & 349.6106 &  -42.3512 & 1.10\\

\enddata
\tablenotetext{a}{Part of the sample of candidate Compton-thick objects in \cite{dellaceca08a}.}
\tablenotetext{b}{NGC 1365 is a complex source that shows a column
density that can vary from LogN$_H\approx23$ to $\geq24$ on timescales
of $\sim$10\,hr \citep{risaliti09a}. According to \cite{risaliti09b}
the source has an absorber with LogN$_H\approx24.6$ which covers $\sim$80\,\%
of the source.}
\tablenotetext{c}{UGC 4203 (also called the `Phoenix' galaxy) is known to exhibit
changes in the absorbing column density from the Compton-thin to the Compton-thick
regime \citep[see e.g.][]{risaliti10}.}
\tablenotetext{d}{Reported to be Compton-thick by \cite{reynolds94}, and \cite{bassani99}.}
\tablenotetext{e}{For the value of the absorbing column density see \cite{dellaceca08a}
and references therein.}
\tablenotetext{f}{\cite{piconcelli11} reports that this source might be absorbed
by LogN$_{\rm H}\geq$25.}
\end{deluxetable}

%
%
\section{Statistical Properties}

\subsection{The Source Count Distribution}
\label{sec:counts}

The source count distribution of radio-quiet AGN (also called  LogN--LogS)
 has already been derived above 10\,keV by several
authors \citep[e.g.][]{ajello08b,tueller08,ajello09b,krivonos10}.
Here we use our larger complete set of AGN to refine the determination
of the  LogN--LogS which is 
of particular
interest since  NuSTAR will be surveying, 
with a factor $>$100 better sensitivity,
a similar energy band in the very near future.
Our aim is to also to compare the BAT observations to the predictions
of popular population synthesis models.

In order to account robustly for uncertainties in the determination
of the LogN--LogS we perform a bootstrap analysis,
creating 1000 resampled set extracted (with replacement) from the BAT dataset.
We perform a maximum likelihood fit \citep[see ][for details]{ajello09b}
to each data set with a power law of the form:
\begin{equation}
\frac{dN}{dS}= A* (S/10^{-11})^{-\alpha}
\end{equation}
where $S$ is the source flux,
and $\alpha$ and $A$ are respectively
the slope and the normalization of the power law.
From the 1000 realizations of the BAT AGN set we derive the distributions
of the normalization and of the slope and we use these to determine
the best-fit parameters and their associated errors.

\begin{figure}[ht!]
  \begin{center}
  	 \includegraphics[scale=0.90]{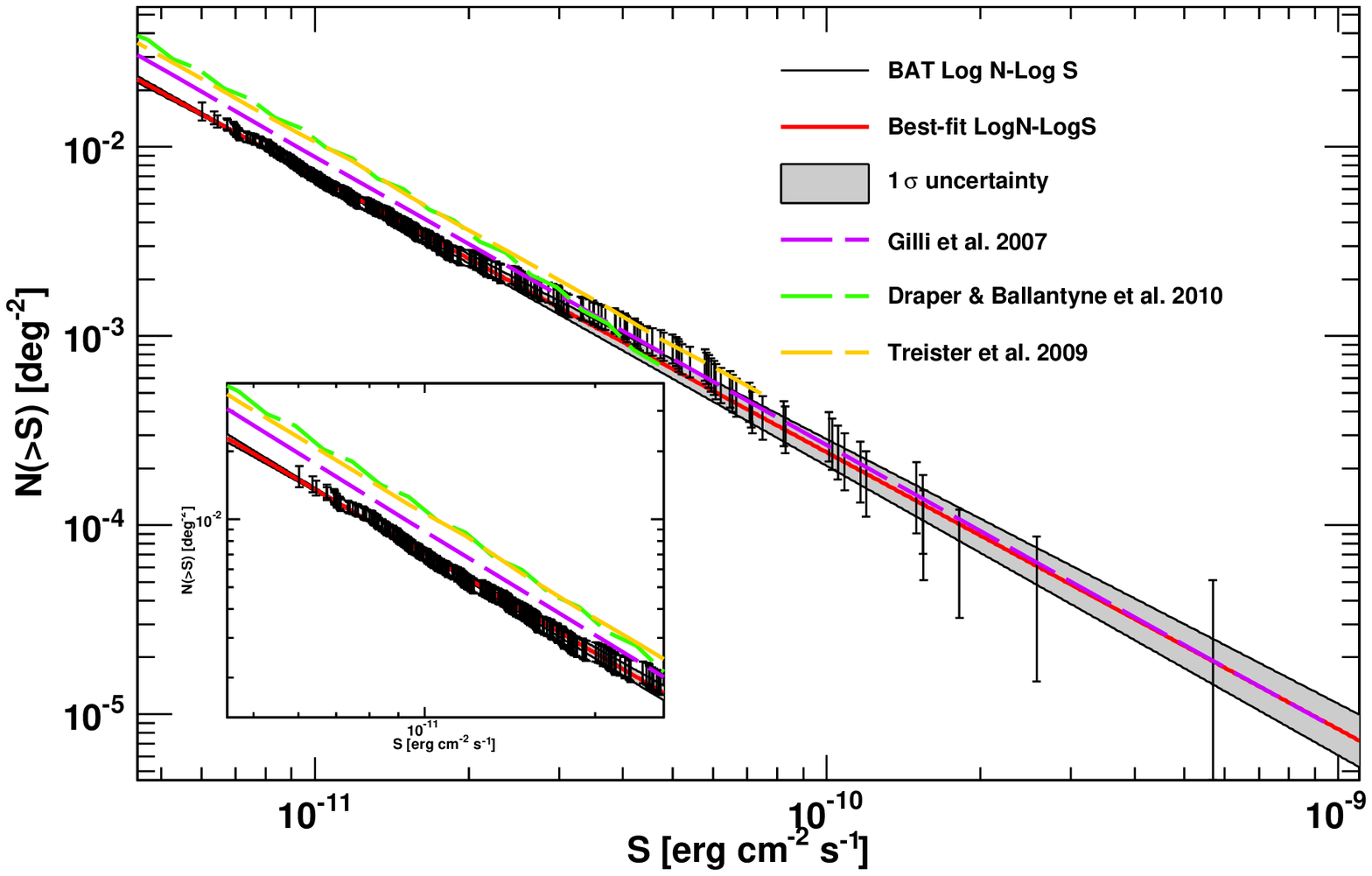}
  \end{center}
  \caption{LogN-LogS of the BAT AGN (black data points) and best power-law fit
in the 15-55\,keV band.
The shaded gray region represents the 1\,$\sigma$ uncertainty computed 
via bootstrap. The dashed lines show the predictions of the number counts
from the models of \cite{gilli07}, \cite{treister09}, and \cite{draper10}.
The inset shows a close-up view of the distribution at the lowest fluxes.
\label{fig:logn}}
\end{figure}

From our analysis we find the best-fit values of: $\alpha$=2.49$^{+0.08}_{-0.07}$
and A=1.05$^{+0.04}_{-0.04}\times10^{9}$.
So the BAT LogN--LogS is compatible with Euclidean for all fluxes spanned by 
this analysis as shown in Fig.~\ref{fig:logn}.
 The surface density of AGN at fluxes (15-55\,keV) grater than 
10$^{-11}$\,erg cm$^{-2}$ s$^{-1}$ is 6.67$^{+0.11}_{-0.12}\times10^{-3}$\,deg$^{-2}$ which is in good agreement with the value of 6.7$\pm0.4\times10^{-3}$\,deg$^{-2}$ reported in \cite{ajello09}.

In order to compare our results with the LogN-LogS measurements
published elsewhere we adopt the following two strategies to convert
fluxes from one band to another. In the first case we adopt
a simple power law with a photon index of 2.0 which is known to describe
generally well the spectra of faint AGN in the BAT band \citep{ajello08b}.
Additionally we use a more complex model for the AGN emission
in the BAT band which is based on the  PEXRAV model of \cite{magdziarz95}.
In \cite{burlon11} the stacking of $\sim$200 AGN spectra 
revealed that the average AGN spectrum 
is curved in the 15--200\,keV band. This stacked AGN spectrum  can be described
 using a  PEXRAV model
with a power-law index of 1.8, an energy cut-off of 300\,keV and
a reflection component due to a medium that covers an angle of
2$\pi$ at the nuclear source (R$\approx$1 in the PEXRAV model).
These  parameters are reported in Tab.~1 of \cite{burlon11}.
To convert fluxes from the 15--55\,keV band to the e.g. 14--195\,keV
band used by \cite{tueller08} the two factors are 2.02 and 1.92 (for
the power law and PEXRAV model respectively). So we consider
the uncertainty related to the flux conversion  to be 
of the $\sim$5\,\% order.

\begin{figure}[ht!]
  \begin{center}
  	 \includegraphics[scale=0.80]{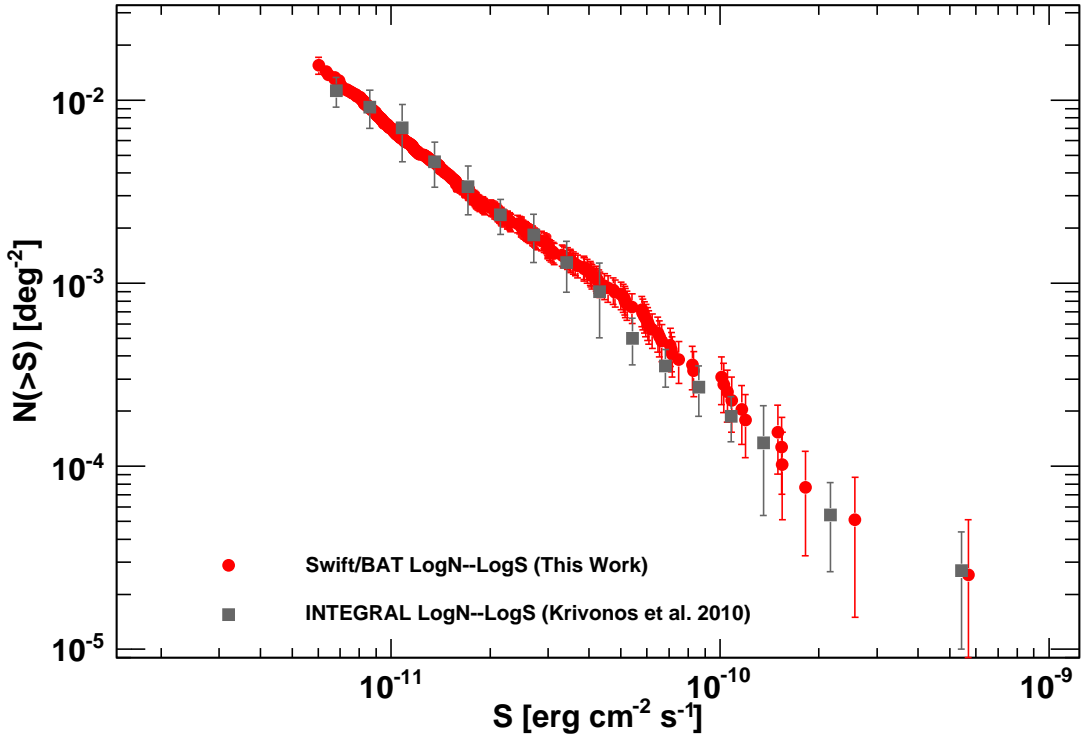}
  \end{center}
  \caption{LogN-LogS of the BAT AGN in the 15--55\,keV band (filled circles) 
compared to the one derived from INTEGRAL data by \cite{krivonos10}
(squares)
\label{fig:logn_krivonos}}
\end{figure}

We compare in Tab.~\ref{tab:logn} our results to those of \cite{tueller08},
\cite{cusumano09} and \cite{krivonos10}. When comparing INTEGRAL and 
BAT results  one has to take into account the
different normalizations of the Crab spectrum that the two instruments
adopt \citep[][]{krivonos10}. To make a proper comparison
we convert the INTEGRAL 17--60\,keV LogN--LogS to the 15--55\,keV BAT band
taking into account the different normalizations\footnote{In order to convert
the INTEGRAL data to the BAT band we took into account that
1\,BAT-mCrab in the 17--60\,keV band is 1.22$\times10^{-11}$\,erg cm$^{-2}$ s$^{-1}$ while 1\,INTEGRAL-mCrab is  1.43$\times10^{-11}$\,erg cm$^{-2}$ s$^{-1}$.
Subsequently we converted the 17--60\,keV fluxes to the 15--55\,keV adopting
a power law with an index of 2.0. This leads to 
F$_{15-55}^{BAT}$=0.878 F$_{17-60}^{INTEGRAL}$.}.
It is apparent that there is 
excellent agreement with the results of \cite{tueller08} both in term
of normalization of the LogN--LogS and also in term of its slope.
Both slopes reported by \cite{cusumano09} and \cite{krivonos10} are in
agreement with ours,  but the density of AGN  reported by \cite{cusumano09}
is smaller than ours. We find an overall agreement within $\sim$10\,\%
of our results and the ones of \cite{krivonos10} as Fig.~\ref{fig:logn_krivonos} testifies.

We also compare the BAT results and the results of the  population synthesis
models of \cite{gilli07}, \cite{treister09} and \cite{draper10}.
For both the Treister et al., and Draper \& Ballantyne models we
used the predictions for the 10--30\,keV band reported in \cite{ballantyne11}
while for the \cite{gilli07} model we use the 10--40\,keV predictions
available online\footnote{The \cite{gilli07} model is available at 
http://www.bo.astro.it/~gilli/counts.html.}. 
Since BAT detects very few Compton-thick AGN we limited  the predictions
of the models of \cite{gilli07} and \cite{draper10} to objects with 
LogN$_{\rm H}\leq$24. The predictions by  \cite{treister09} include objects
with LogN$_{\rm H}\geq$24, however in their modeling the density of Compton-thick
AGN is (at BAT sensitivities) $\sim$7\,\% of the total AGN population.
The predictions
of all models, converted to the BAT band using the above prescriptions,
are compared to the BAT LogN--LogS in Fig.~\ref{fig:logn}.
It is apparent that there is good agreement with the BAT results 
at bright fluxes (i.e. $>10^{-11}$\,erg cm$^{-2}$ s$^{-1}$).
At the limiting flux of our analysis (i.e. $\sim$6$\times10^{-12}$\,erg cm$^{-2}$ s$^{-1}$) the model predictions are beyond the statistical uncertainty
of the BAT LogN--LogS as shown in Tab.~\ref{tab:logn_model} and in
the inset of Fig.~\ref{fig:logn}.

The model of \cite{gilli07} is compatible at bright fluxes with
the BAT data, but with a steeper slope. On the other hand
there seems to be a constant offset between the BAT LogN-LogS and
the model predictions of \cite{treister09} and \cite{draper10}.
The model of \cite{treister09} reproduces  the 14--195\,keV 
BAT LogN--LogS of \cite{tueller08} which is in very good agreement
with the one published here. However, a close inspection 
\citep[see Fig.~1 in ][]{treister09} shows that the \cite{tueller08}
data (reported in \citealt{treister09}) 
have a normalization $\sim$1.6 larger than the original measurement
reported in \cite{tueller08}. Thus the \cite{treister09} is anchored
to LogN-LogS data with a normalization $\sim$1.6 larger than
observed. This same factor is apparent when comparing the \cite{treister09}
prediction to our data (see Tab.~\ref{tab:logn_model}).

The model of \cite{draper10} reproduces (see their Fig.~3)
the `correct' 14--195\,keV LogN--LogS as reported by \cite{tueller08}.
However,  their 10--30\,keV model prediction is at fluxes $\geq10^{-14}$\,erg
cm$^{-2}$ s$^{-1}$ very similar to the one of \cite{treister09}
and their predicted densities are a factor 1.6--1.8
larger than the measured ones  in the 15--55\,keV band
(see Tab.~\ref{tab:logn_model}) at the faintest
fluxes sampled by our analysis.
These findings might have some implications for the number of
objects predicted to be detected by NuSTAR in serendipitous
surveys above 10\,keV (see $\S$~\ref{sec:nustar}).
Given the substantial agreement of the BAT and INTEGRAL LogN--LogS
(see Fig.~\ref{fig:logn_krivonos}), it does not seem
 likely that the discrepancy in the predictions of synthesis
models and the $>$10\,keV LogN--LogS can be ascribed to 
a difference in the results above 10\,keV.

\begin{deluxetable}{lcc}
\tablewidth{0pt}
\tablecaption{Properties of LogN--LogS derived above 10\,keV
\label{tab:logn}}
\tablehead{ \colhead{Model} & 
\colhead{Surface Density\tablenotemark{a}}  & \colhead{$\beta$} \\
\colhead{} & \colhead{\scriptsize (10$^{-3}$\,deg$^{-2}$)} & \colhead{}
}
\startdata
This Work & 6.67$^{+0.11}_{-0.11}$ & 2.49$^{+0.08}_{-0.07}$\\
Tueller et al. 2008 & 6.5--6.8& 2.42$\pm0.14$\\
Cusumano et al. 2009 & 5.4--5.9 & 2.56$\pm0.06$\\
Krivonos et al. 2010 & 7.0--8.1 & 2.56$\pm0.10$\\

\enddata
\tablenotetext{a}{Density at a 15-55\,keV flux of 
10$^{-11}$\,erg cm$^{-2}$ s$^{-1}$}
\end{deluxetable}

\begin{deluxetable}{lc}
\tablewidth{0pt}
\tablecaption{Comparison of the BAT LogN--LogS with synthesis models.
\label{tab:logn_model}}
\tablehead{ \colhead{Model} & 
\colhead{Surface Density\tablenotemark{a}}\\
\colhead{} & \colhead{\scriptsize (10$^{-2}$\,deg$^{-2}$)}
}
\startdata
This Work                              & 1.49$^{+0.04}_{-0.03}$ \\
Gilli et al. 2007\tablenotemark{b}     & 2.14--2.35   \\
Treister et al. 2009\tablenotemark{c}  & 2.31--3.02  \\
Draper \& Ballantyne 2010              & 2.52--3.20 \\

\enddata
\tablenotetext{a}{Density at a 15-55\,keV flux of 
6$\times^{-12}$\,erg cm$^{-2}$ s$^{-1}$.}
\tablenotetext{b}{To convert the original 10--40\,keV counts
to the BAT 15--55\,keV band we have used the following
factors: 0.94 and 1.04 for the power law and PEXRAV model.}
\tablenotetext{c}{To convert the  10--30\,keV counts 
\citep[reported in ][]{ballantyne11}
to the BAT 15--55\,keV band we have used the following
factors: 1.18 and 1.40 for the power law and PEXRAV model.}

\end{deluxetable}

%
%
\subsection{The Luminosity Function}
\label{sec:xlf}

X-ray selected (below 10\,keV)
AGN are known to display a luminosity function that evolves with
redshift \cite[see e.g.][]{ueda03,lafranca05,hasinger05,silverman08,aird10}.
Our sample reaches a redshift of z$\approx$0.3 where 
according to the above works, the evolution of AGN is significant and
can potentially be detected.
In \cite{ajello09b} we found marginal evidence for the evolution
of AGN in the local Universe. Here we can make use of our all-sky sample
of Seyfert galaxies to test this hypothesis.    
We adopt as a description of the X-ray luminosity function (XLF) a pure luminosity evolution model as follows:
\begin{equation}
\frac{dN}{dVdL_X}= \Phi(L_X(z),z)=\Phi(L_X/e(z)) 
\label{eq:ple}
\end{equation}
where $V$ is comoving volume element and the 
evolution is parametrized using the common  power-law evolutionary  factor:
\begin{equation}
e(z)= (1+z)^k
\label{eq:ez}
\end{equation}

We neglect any cut-off in the evolution as this takes place at a redshift
that BAT cannot constrain \citep[i.e. z$\approx$1 see e.g.][]{aird10}.
For the XLF  at redshift zero we use a  double power-law of the form:
\begin{equation}
\Phi(L_X,z=0) = \frac{dN}{dL_X}=
\frac{A}{ln(10)L_X}\left[ \left(\frac{L_X}{L_*} \right) ^{\gamma_1} 
+ \left(\frac{L_X}{L_*}\right)^{\gamma_2} \right]^{-1}
\label{eq:2pow}
\end{equation}

A value of $k$ significantly different than zero would point towards 
an evolution of the XLF.
In order to the derive the luminosity
function of AGN, we adopt the same maximum likelihood method described in 
\cite{ajello09b}. In particular we determine the best-fit parameters of the XLF
by finding the minimum  of Eq.~11 in \cite{ajello09b}.

In order not to include sources which could have a non negligible
contribution to their total luminosity 
from X-ray binaries \cite[see e.g.][]{voss10}
and to limit the incompleteness due to the bias against the detection
of the most absorbed sources we derive the XLF only for LogL$_{X}\geq$41.3.
The best-fit parameters for the PLE model in
case of evolution ($k\neq$0) and in case of no evolution
($k=$0) are reported in Tab.~\ref{tab:pars}.
The first result is that 
the model with no evolution (i.e. Eq.~\ref{eq:2pow})
represents a good  description of the BAT dataset.
The best-fit parameters are compatible with those reported by 
\cite{sazonov08}, \cite{tueller08} and \cite{ajello09b}.

Adding the extra $k$ parameter produces only a marginal improvement
in the fit. Indeed the log-likelihood improves only by $\sim$5 which
corresponds to a $\sim$2\,$\sigma$ improvement \citep{wilks38}.
This is reflected in the luminosity evolution parameter 
which is constrained to be  $k$=1.38$\pm0.61$.
The value found here is compatible, but smaller than the value of 2.62$\pm1.18$
reported in \cite{ajello09b} showing that if there is evolution
in the XLF of local AGN that might be shallower than previously found.
We thus believe that the non-evolving XLF model is, in view of 
the marginal improvement of the goodness of fit, a better representation
of the current dataset. Nevertheless, we will use the PLE model
in $\S$~\ref{sec:nustar} to assess the level of uncertainty in the
prediction for the number of  AGN that might be detected by NuSTAR
in a very near future.

It is interesting to compare our XLF with that one reported by \cite{sazonov08}
and \cite{tueller08} in the 17--60\,keV and in the 14--195\,keV band respectively.
\cite{sazonov08} and \cite{tueller08} report a value of the faint-end
slope $\gamma_1$ of 0.76$^{+0.18}_{-0.20}$ and 0.84$^{+0.16}_{-0.22}$ 
respectively. The value of our faint-end slope is 0.78$\pm0.08$ in 
agreement with both results, but much better constrained because of
the larger dataset. When converted to our band (see previous section
for the conversion factors), the break luminosity of  
\cite{sazonov08} and \cite{tueller08} is respectively 
2.2$^{+2.0}_{-1.0}\times10^{43}$\,erg s$^{-1}$ 
and  3.7$^{+3.0}_{-1.6}\times10^{43}$\,erg s$^{-1}$ while we measure
5.1$\pm1.4\times10^{43}$\,erg s$^{-1}$ again compatible, but better constrained.

In order to display the LF we rely on the ``N$^{obs}$/N$^{mdl}$'' method
devised by \cite{lafranca97} and \cite{miyaji01} and 
employed in several recent works \citep[e.g.][]{lafranca05,hasinger05}.
Once a best-fit function for the LF has been found, it is possible
to determine the value of the observed LF in a given bin of luminosity
and redshift:
\begin{equation}
\Phi(L_{X,i},z_i) = \Phi^{mdl}(L_{X,i},z_i) \frac {N^{obs}_i}{N^{mdl}_i}
\end{equation}
where $L_{X,i}$ and $z_i$ are the luminosity and redshift of the i$^{th}$
bin, $\Phi^{mdl}(L_{X,i},z_i)$ is the best-fit LF model and $N^{obs}_i$ 
and $N^{mdl}_i$ are the observed and the predicted number of AGN in that bin.

\begin{figure}[ht!]
  \begin{center}
  	 \includegraphics[scale=0.80]{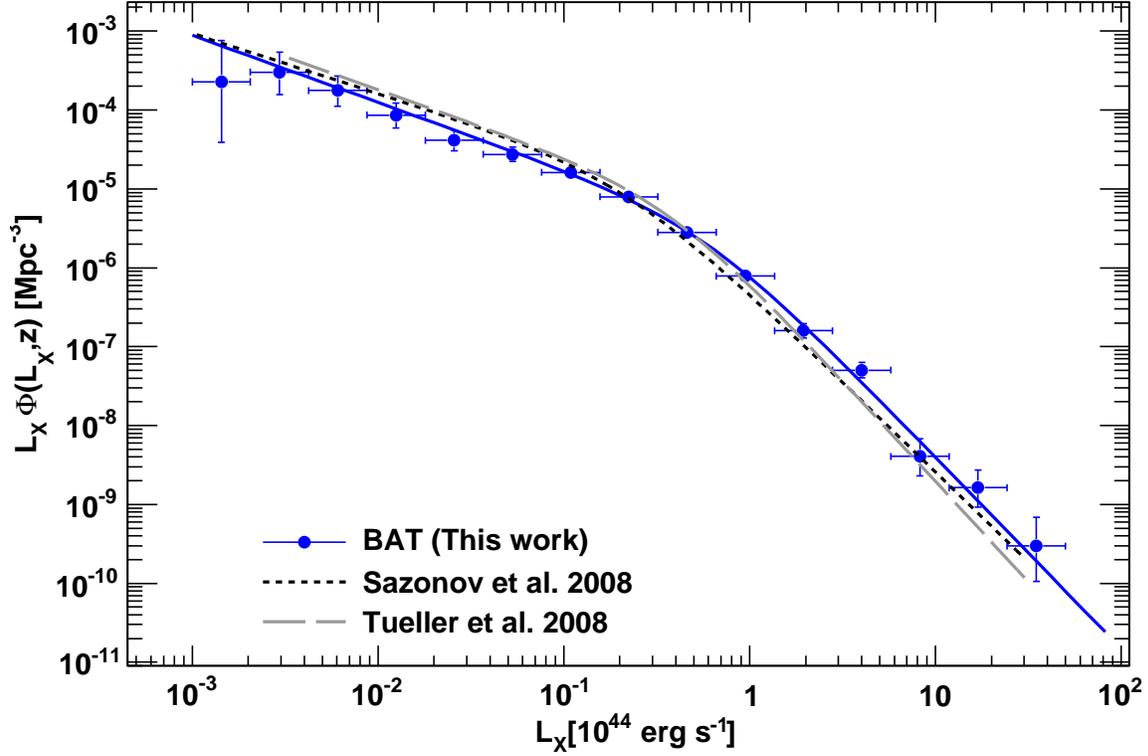}
  \end{center}
  \caption{Luminosity function in the 15--55\,keV band
of the BAT AGN (non evolving model) 
compared with the measurements of \cite{tueller08} and \cite{sazonov08}.
The data point at a luminosity $<2\times10^{41}$\,erg s$^{-1}$
was not fitted to avoid problems related to incompleteness in 
detecting Compton-thick AGN and to avoid contamination from
sources whose flux might be dominated by the  emission
of X-ray binaries \citep{voss10}.
\label{fig:xlf}}
\end{figure}

Fig.~\ref{fig:xlf} shows the best-fit non-evolving model (i.e. $k$=0)
in comparison with the XLF of \cite{sazonov08} and \cite{tueller08}.
In general there is very good agreement between the XLFs derived in all
these works. 
Our results are also in agreement with those obtained in the $<$10\,keV band.
Indeed, for the bright end slope \cite{barger05}, \cite{lafranca05} and
\cite{aird10} obtain 2.2$\pm0.5$, 2.36$^{+0.13}_{-0.11}$ and 2.55$\pm0.12$ 
respectively while we measure 2.39$\pm0.12$.
For the faint-end slope \cite{barger05}, \cite{lafranca05} and \cite{aird10}
report 0.42$\pm0.06$, 0.97$^{+0.08}_{-0.07}$ and 0.58$\pm0.04$\footnote{\cite{aird10} reports also 0.70$\pm0.03$ for their PLE model.} while we measure
$\gamma_1=$0.78$\pm0.08$. The agreement for the faint-end slope is not
as a good as for the bright-end slope and there is some scatter (that
appears to be systematic in origin) in
the $<$10\,keV measurement while there is substantial agreement above 10\,keV
(see above discussion). An excessive flatness of the faint-end slope
might be linked to the role of Compton-thick AGN which are difficult
to detect because the absorption pushes their observed luminosity
below the survey threshold. Indeed, as shown
in \cite{burlon11} 2--10\,keV surveys are more biased in the detection
of Compton-thick AGN than surveys above 10\,keV. Depending on the redshift
distribution of the sources (and thus on the k-correction) the bias
might not be the same for different 2--10\,keV surveys.

For completeness we also report in Tab.~\ref{tab:pars}
the best-fit parameters to the XLF
of all AGN excluding the Compton-thick ones of Tab.\ref{tab:ct}.
Because Compton-thick objects represent a small fraction of the AGN
detected by BAT, there is very little difference between the XLF 
of all AGN and that of AGN with  LogN$_{H}<$24. 
However, in the next section this XLF  will be useful to compare
the space density of Compton-thick AGN to that of Compton-thin (LogN$_{H}<$24)
 AGN.

\begin{deluxetable}{llcccc}
\tablewidth{0pt}
\tablecaption{Parameters of fitted Luminosity Functions. Parameters without an
error estimate were kept fixed during the fitting stage.
\label{tab:pars}}
\tablehead{ \colhead{Model} & \colhead{A\tablenotemark{a}} & \colhead{$\gamma_1$} &
\colhead{$\gamma_2$} & \colhead{$L_*$\tablenotemark{b}} & \colhead{k} 
}
\startdata
No evolution & 113.1$\pm6.0$\tablenotemark{a} & 0.79$\pm0.08$ & 2.39$\pm0.12$ & 0.51$\pm0.14$ & 0\\

PLE  & 109.6$\pm5.3$\tablenotemark{a} & 0.78$\pm0.07$ & 2.60$\pm0.20$ & 0.49$\pm0.10$ &1.38$\pm0.61$ \\

No evol., no CT & 122.4$\pm6.6$\tablenotemark{a} & 0.72$\pm0.09$ & 2.37$\pm0.12$ & 0.48$\pm0.13$ & 0\\

\enddata
\tablenotetext{a}{In unit of $10^{-7}$\,Mpc$^{-3}$.}
\tablenotetext{b}{In unit of $10^{44}$\,erg s$^{-1}$.}
\end{deluxetable}

%
%
%
\subsection{The Space Density of Compton-thick AGN}
\label{sec:cthick_dens}

We derive the space density of Compton-thick AGN
with 24$\leq$LogN$_{\rm H}\leq$25 using
the 1/V$_{\rm MAX}$ non-parametric method \citep{schmidt68}.
We de-absorb the luminosities
of the BAT Compton-thick sources reported in Tab.~\ref{tab:ct}
using the correction function reported in \cite{burlon11} (see their Fig.~11).
This correction function was derived for the average properties of the
Compton-thick AGN detected by BAT and takes into account photoelectric absorption
as well as Compton scattering \citep[see][for details]{murphy09}.

\begin{figure}[ht!]
  \begin{center}
  	 \includegraphics[scale=0.80]{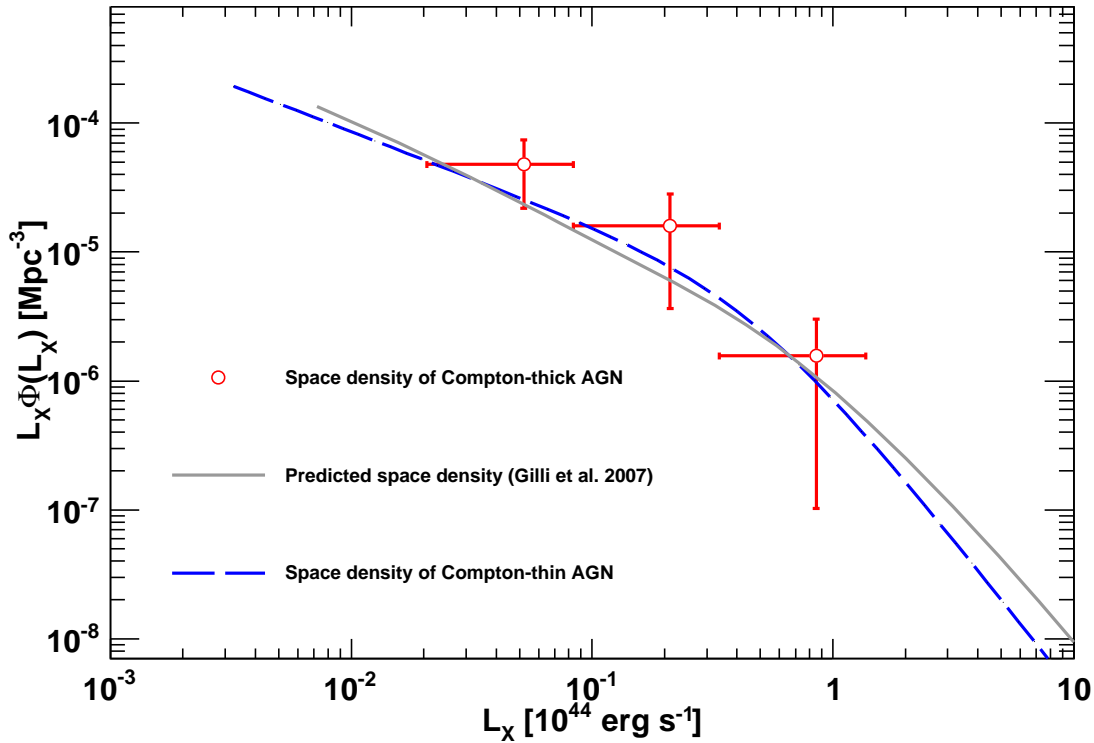}
  \end{center}
  \caption{Space densities of Compton-thick AGN detected by BAT compared
to model prediction from \cite{gilli07} and the space density of Compton-thin
(LogN$_{\rm H}<24$) AGN
from $\S$~\ref{sec:xlf}. Luminosities were de-absorbed following the method
outlined in \cite{burlon11}.
\label{fig:ct_sp}}
\end{figure}

\begin{figure}[ht!]
  \begin{center}
  	 \includegraphics[scale=0.80]{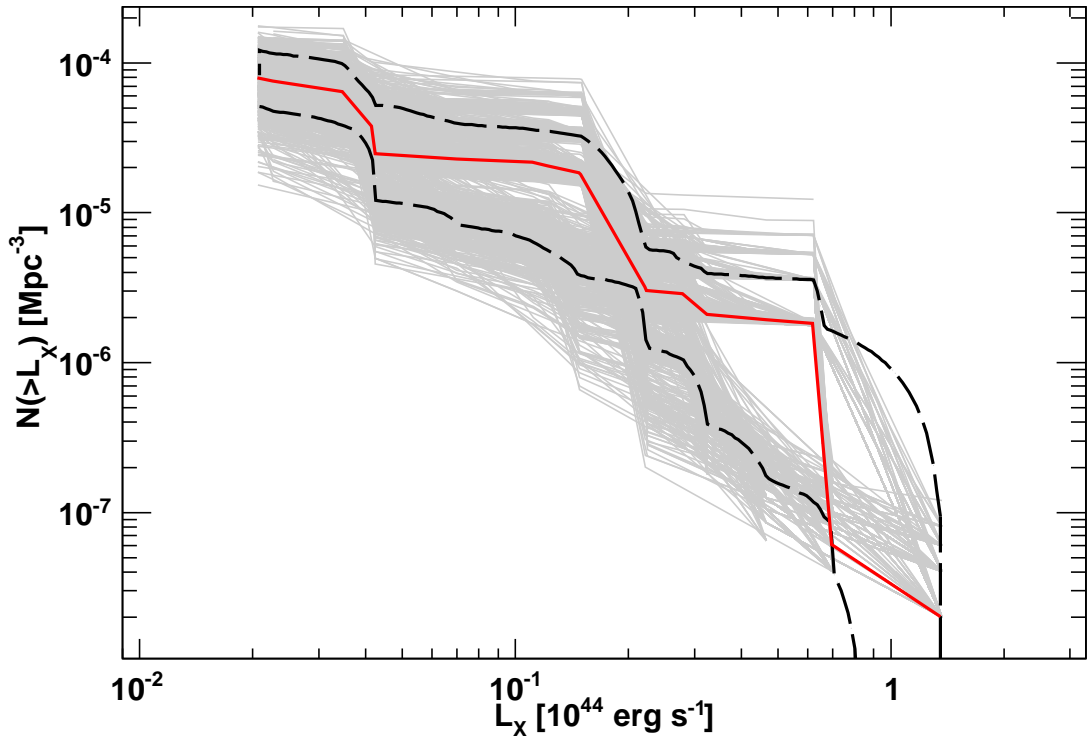}
  \end{center}
  \caption{Cumulative space densities of Compton-thick AGN detected by BAT
(thick red solid line) with 1\,$\sigma$ confidence contours (dashed line)
generated from the analysis of the bootstrapped samples (thin gray lines).
\label{fig:ct_cum}}
\end{figure}

Fig.~\ref{fig:ct_sp} shows the luminosity function of the Compton-thick AGN detected by BAT while Fig.~\ref{fig:ct_cum} reports the cumulative  space density.
In this latter case the uncertainty were computed via bootstrap with replacement.
We find that the space density of Compton-thick AGN with
a de-absorbed luminosity greater than 2$\times10^{42}$\,erg s$^{-1}$ is
7.9$^{+4.1}_{-2.9}\times10^{-5}$\,Mpc$^{-3}$. Above a de-absorbed luminosity of 
$10^{43}$\,erg s$^{-1}$
the density becomes 2.1$^{+1.6}_{-1.4}\times10^{-5}$\,Mpc$^{-3}$.
As shown in Fig.~\ref{fig:ct_sp}  the model predictions of \cite{gilli07}
are compatible within the statistical error with our space density estimates.
\cite{treister09} and \cite{draper10} estimated a space density of 
LogL$_{X}\geq$43\,erg s$^{-1}$ sources of 
respectively  
2.2$^{+2.9}_{-1.1}\times10^{-6}$\,Mpc$^{-3}$  and
3--7$\times10^{-6}$\,Mpc$^{-3}$. These estimate are below
ours, but compatible within 2$\sigma$ and 1$\sigma$ respectively.

In the same Fig.~\ref{fig:ct_sp} we plot also  the space density of AGN 
with LogN$_{\rm H}<$24 (from the previous section). 
While there is  substantial agreement between the two, our analysis
seems to point to the fact that the space density of Compton-thick AGN
is larger than the one of all other classes of AGN. 
By allowing the normalization $A$
of the XLF of LogN$_{\rm H}<$24 AGN to vary we find 
that $A_{\rm LogN_{H}\geq24}$=1.4$\times A_{\rm LogN_{H}<24}$: i.e.
 the space density of Compton-thick AGN is 1.4 times larger than that of
LogN$_{\rm H}<$24 AGN.  A similar results was obtained by 
\cite{dellaceca08b} who derived indirectly the space density of Compton-thick AGN
as a difference between the density of optically selected AGN and that of
X-ray selected AGN with LogN$_{\rm H}<$24 (see the aforementioned paper for more
details). In their study they also 
find that the space density of  Compton-thick AGN 
with LogL$_{X}\geq$43\,erg s$^{-1}$ is $\sim$1.6$\times10^{-5}$\,Mpc$^{-3}$
which is in good agreement with the value found here.

%
%
%
\section{The Anisotropic Local Universe}
\label{sec:aniso}

The BAT sample represents an incredible resource to study the properties
of the nearby Universe as it is a truly local sample of AGN.
Indeed, the median redshift of the radio-quiet AGN detected by BAT is $\sim$0.03
which corresponds to $\sim$120\,Mpc. On these scales the
structure of the local Universe is known to be inhomogeneous and
AGN are known to trace the matter density distribution 
\citep{cappelluti10}. The BAT AGN can thus be used to trace the large-local
scale structure.
\cite{krivonos07} reported, using 68 AGN
detected by INTEGRAL, an anisotropy of the spatial distribution of local AGN.
With an AGN sample $\sim$6 times larger than the INTEGRAL sample it is
possible to identify the anisotropic spatial distribution of AGN
to a higher precision than previously obtained.

The following approach is adopted in order to assess the anisotropy of the 
spatial distribution of AGN. We first separate the sample in two redshift
bins z$<$0.02 and 0.02$\leq$z$\leq$0.15 corresponding to a distance
smaller (or grater) than 85\,Mpc. The choice is dictated by the fact
that the largest contrast in AGN density is expected to be observed 
nearby from us.

For each of these sub-samples we create a LogN--LogS that yields
the average surface density of AGN for that redshift slice.
Taking into account how the BAT sensitivity varies across the sky
we generate (for each redshift bin) 1000 realizations of AGN sets
which are  isotropically distributed in the sky.
Then  for each direction in the sky we count how many AGN BAT has detected 
within a radius of 20$^{\circ}$ (typically this number
oscillates around 10) versus  the expected number for the
isotropic case. For each of these sky positions we compute
the fractional over-density of AGN: i.e. the ratio between the number
of detected AGN and the number of expected objects if AGN were isotropically
distributed. From the 1000 realizations we also compute the error (and 
the significance) connected to the fractional over-density.

\begin{figure*}[ht!]
  \begin{center}
\vspace{-0.5cm}
  \begin{tabular}{c}
    \includegraphics[scale=0.8]{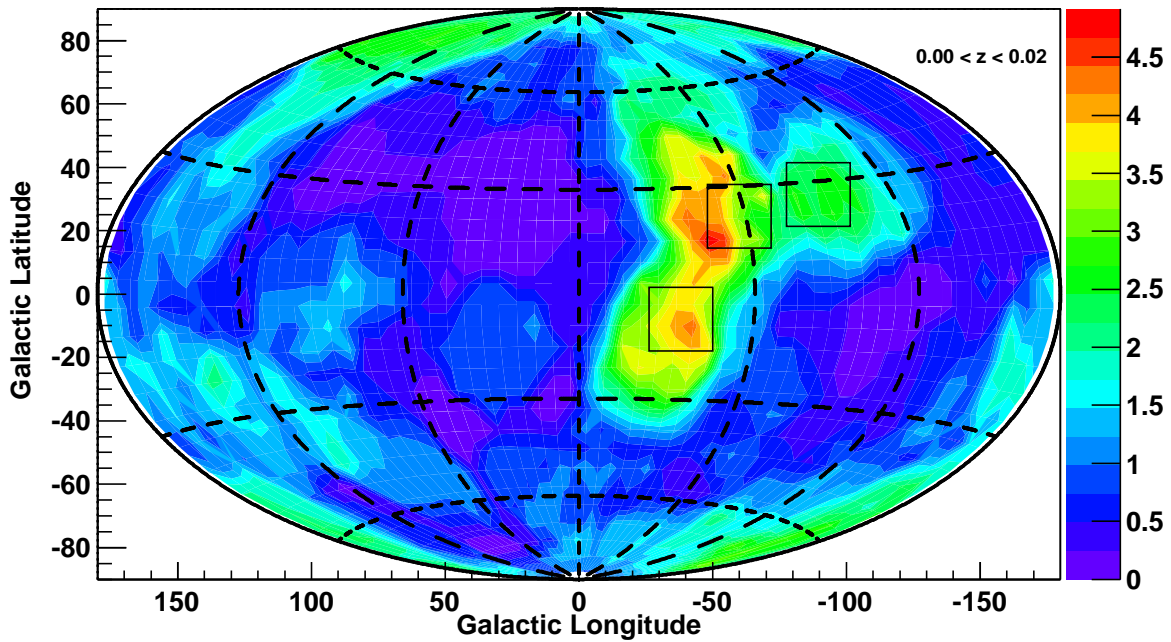}\\
\end{tabular}
  \end{center}
  \caption{Fractional over-density of AGN 
with respect to the isotropic expectation  within 85\,Mpc.
The color scale shows the ratio of BAT-detected AGN (within 20\,degree cones)
to the average number of AGN expected in the same area if sources
were isotropically distributed.
The map shows the regions where this ratio is the largest (red)
and where it is the smallest (blue). The squares show the approximate
position of the most prominent super-clusters (see text for details).
\label{fig:aniso}}
\end{figure*}

\begin{figure*}[ht!]
  \begin{center}
  \begin{tabular}{c}
    \includegraphics[scale=0.8]{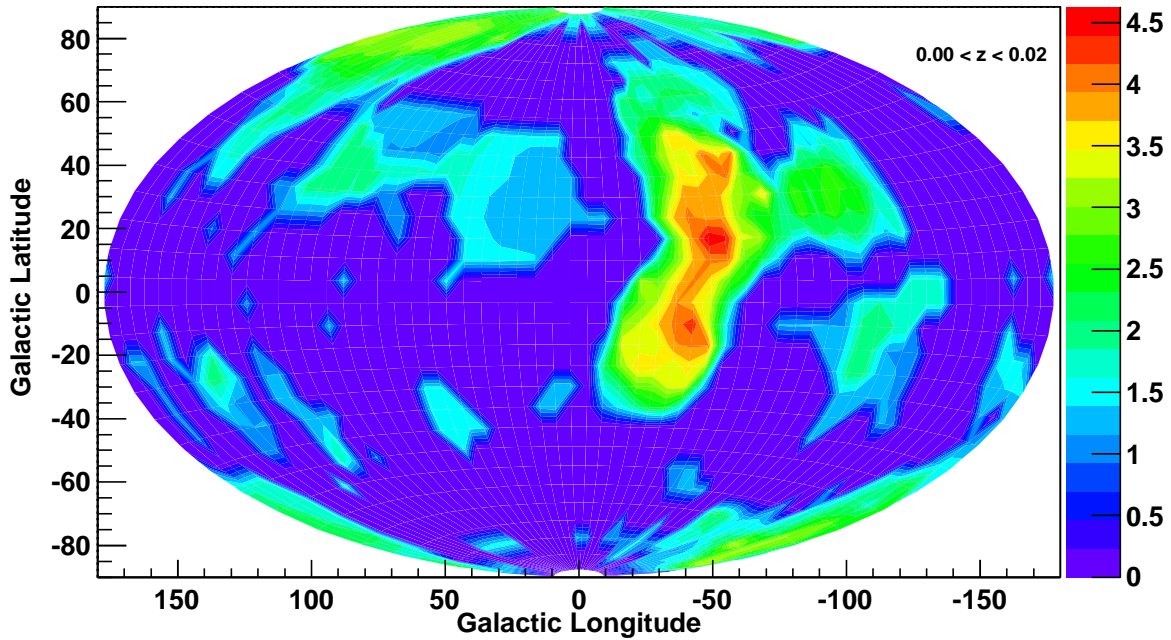}\\
\end{tabular}
  \end{center}
  \caption{
Significance of the density features of Fig.~\ref{fig:aniso}
expressed in number of $\sigma$.
\label{fig:snr}}
\end{figure*}

Fig.~\ref{fig:aniso} and Fig.~\ref{fig:snr}  
show how the fractional over-density of  
AGN (i.e. the ratio of detected AGN to the expected 
number for the isotropic case) and its significance
change across the sky for the lower redshift bin. 
It is apparent that  there is a marked contrast in the AGN density
within 85\,Mpc, with the AGN density varying  by a factor of $>$10
across the sky. The most significant over-density is seen in the first
redshift slice (i.e. within 85\,Mpc)  with prominent structures significant
at $>$4\,$\sigma$.

There is a clear over-density of AGN (a factor $>$4 larger than
the isotropic expectation) in the direction of the super-galactic plane. 
In particular the most dense region
is found to be at $l=-54^{\circ}$  extending from $b=\sim15^{\circ}$ to 
$b=\sim50^{\circ}$. This
corresponds to the position of the  Hydra-Centaurus super-cluster (z=0.01-0.02)
which represents one of the largest structures in the local Universe.
The second most dense region (connecting to the south 
of the Hydra-Centaurus super-cluster) can likely be identified with the 
`Great Attractor' \citep{lynden88}. 
In Fig.~\ref{fig:aniso} the positions of the
Great Attractor, the Centaurus and Hydra superclusters are shown
with squares (from bottom left to upper right, respectively).
The redshifts of the BAT AGN in these dense regions are in agreement
with the redshifts of these massive structures.
Our results appear to be in good agreement with the ones of \cite{krivonos07},
but the improved statistics allow us to locate more  precisely the over-density
of AGN in the nearby Universe.

In the 0.02$\leq$z$\leq$0.15 the anisotropy is less pronounced and less
significant as well, with the most significant structures being $<3$\,$\sigma$.
A comparison with our simulations (which allow us to account for the trial
factor) shows that the 0.02$\leq$z$\leq$0.15 over-density map 
is indistinguishable  (with the present data) from the isotropic expectation.
However, 
the local environment is known to be highly inhomogeneous up to (at least) 
z$\approx$0.05 \citep{jarrett04}. The reason why BAT does not detect
this anisotropy in the local scale structure is ultimately due to statistics.
Indeed, the typical clustering length of the BAT AGN is known to be 5-8\,Mpc
\citep{cappelluti10}. Within 85\,Mpc, our 20\,degree search cone subtends\footnote{The average redshift of the z$<$0.02 sample is 0.01.}
a length of $\sim$16\,Mpc which is comparable to the clustering length
of AGN. However, for the average redshift ($\sim$0.05) 
of the 0.02$\leq$z$\leq$0.15 sample, the 20\,degrees search cone subtends
a length of 70\,Mpc which is ten times larger than the typical clustering
length of AGN. This contributes to dilute any over-density signal.
In order to have a similar resolution as for the $<$85\,Mpc sample, one
should adopt a search cone of 5\,degrees. However, with the current statistics
we would expect less than 1\,AGN in such cone.

\begin{figure}[ht!]
  \begin{center}
  	 \includegraphics[scale=0.80]{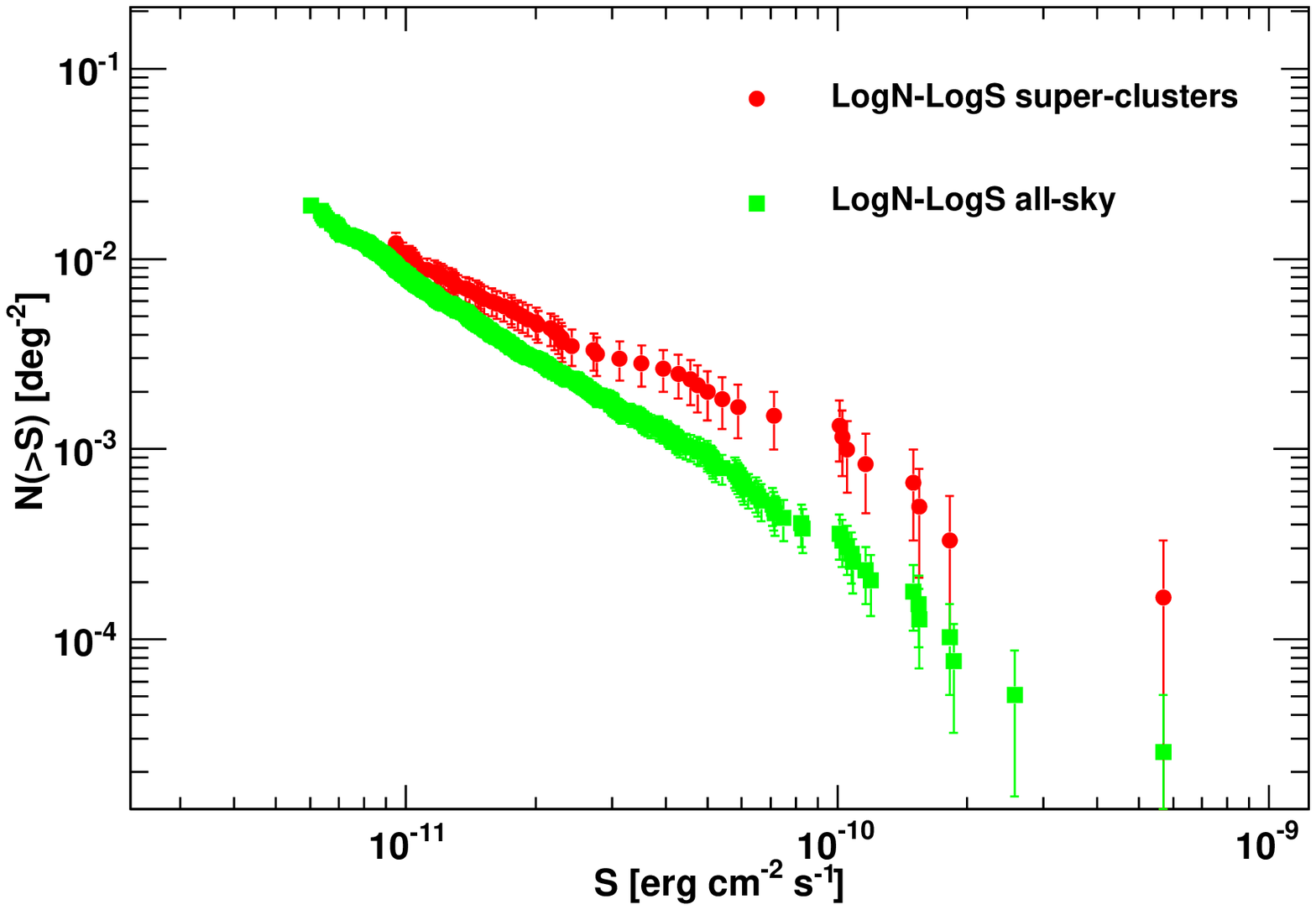}
  \end{center}
  \caption{LogN--LogS (15--55\,keV) of the AGN in the direction of the Hydra-Centaurus
super-clusters and of the Great Attractor compared to the LogN--LogS of all
AGN.
\label{fig:logn_hydra}}
\end{figure}

Finally we checked if the anisotropy depends on any of the AGN parameters
(i.e. flux, luminosity, type, etc.). To this extent we isolated the AGN
in the direction of the Hydra-Centaurus super-clusters
and of the Great Attractor and created a LogN--LogS and a luminosity function.
The first result is that the XLF of these AGN is in good agreement with
that of the entire population reported in Fig.~\ref{fig:xlf}.
The LogN-LogS of the AGN in the direction of the super-clusters is shown,
in comparison with the LogN-LogS of the whole population,
in Fig.~\ref{fig:logn_hydra}. It is clear that there is an excess at bright fluxes. This finding in conjunction with the fact that the XLF does not change
indicates that the sources that contribute to the over-density are the brightest sources
(i.e. the proximity of these local structures makes these sources appear
with a bright flux). Indeed, the median redshift of the AGN in the direction
of the super-clusters (z$\approx$0.015) 
is markedly smaller than that of the whole population (z$\approx$0.03).

It is also interesting to note that the fraction of Sy2 galaxies in the direction
of the super-clusters is larger than average. Indeed, the fraction of objects
classified as Sy2 in our whole sample and in the sample of \cite{cusumano10}
is $\sim$34\,\% of the total AGN population. This ratio would increase to 
$\sim$45\,\% if all the objects with no opictal classification reported in this
or the \cite{cusumano10} sample would be in reality Sy2s. 
While it is reasonable to expect that more than 50\,\% of the objects
lacking an optical classification are Sy2 galaxies, it is unlikely that
all of them are Sy2s. So this represents an extreme scenario.

The fraction of Sy2 object  becomes $\sim$50\,\%
if we consider only the AGN in the direction of the super-clusters
and increases to 58$\pm8$\,\% if we restrict to sources within 85\,Mpc.
This seems in agreement
with what reported by \cite{petrosian82} that Sy1s
are more often found to be isolated than Sy2s.
In dense environments encounters between galaxies produce
gravitational interactions
that can trigger gas inflow towards the central
hole and produce AGN activity. 
Galaxy merging is known to provide an efficient way to 
funnel large amount of gas and dust to the 
central black hole \citep[e.g.][]{kauffmann00,wyithe03,croton06}.
Recently, \cite{koss10} found that 24\,\% of all the hosts of the BAT selected
AGN have a companion galaxy within 30\,kpc. This suggests that for a fraction
of moderate luminosity AGN merging is a viable triggering mechanisms.
However, in dense environments major merging is 
not the only process that might be at work. 
Indeed, 
galaxy harassments (i.e. high speed encounters between galaxies) 
can also drive most of the galaxy's gas to the inner 500\,pc \citep{lake98}
and thus trigger AGN activity.
Also minor merging, where the ratio of the masses of the merging galaxies is $>$3
(and probably around $\sim$10), can  lead to Seyfert level of accretion
\citep{hopkins09}. With large quantities of gas available around the hole,
in the early phase of accretion the AGN might have a higher probability 
of being identified as a type-2 AGN.
This might explain the over-abundance of Sy2 galaxies in dense environments.

%
%

\section{Predictions for NuSTAR}
\label{sec:nustar}
In this section we provide predictions for the number of objects
that NuSTAR might see in different types of blank-field surveys.
The predicted number counts are obtained by extrapolating the BAT
LogN--LogS of $\S$~\ref{sec:counts} to lower fluxes under the assumption
that the slope of the LogN--LogS does not change.
Indeed it is reasonable to expect so since there 
are strong indications both from observations in the 2--10\,keV band and from
modeling that the source count distribution breaks only at fluxes 
$\leq10^{-14}$\,erg cm$^{-2}$ s$^{-1}$. 

The source count distribution of AGN is very well determined
in the 2--10\,keV band\footnote{For a typical AGN 
spectrum the 2--10\,keV flux is on average 20\,\% larger than the 15--55\,keV flux.} 
down to fluxes of $<10^{-15}$\,erg cm$^{-2}$ s$^{-1}$
\citep[see e.g.][]{rosati02,cappelluti07,xue11}.
Its slope is known
to be Euclidean down to fluxes of $\sim10^{-14}$\,erg cm$^{-2}$ s$^{-1}$.
For example \cite{cappelluti07} reports a slope of 2.43$\pm0.10$ is good agreement
with our results. A similar result is found by \cite{rosati02}.
Also population synthesis models predict a break in the LogN-LogS at 
$<10^{-14}$\,erg cm$^{-2}$ s$^{-1}$ \citep{gilli07,treister09,draper10}.

The extrapolation to fluxes a factor $>$50 fainter than those
sample by BAT necessarily introduces some uncertainty related to neglecting
the evolution of AGN. This uncertainty will be gauged later on in this section.
However,   the fact that BAT and NuSTAR sample (almost) the same
energy band removes other sources of uncertainties.
NuSTAR will likely perform three different types of surveys: a shallow, a medium and 
a deep survey. The shallow survey, performed combining short (5--10\,ks) exposures,
 might extend over $\sim$3\,deg$^{2}$ reaching
fluxes\footnote{Typical fluxes for NuSTAR are quoted for the 10--30\,keV band. Here we have converted those fluxes to the 15--55\,keV band adopting 
a power law with a photon index of 2.0.} 
$\sim10^{-13}$\,erg cm$^{-2}$ s$^{-1}$. 
The medium survey will likely cover $\sim$1\,deg$^{2}$ with $\sim$50\,ks
pointings reaching a 10--30\,keV flux of $\sim5\times10^{-14}$\,erg cm$^{-2}$ s$^{-1}$ while
the deep survey is expected to reach a 10--30\,keV flux of 
$\sim2\times10^{-14}$\,erg cm$^{-2}$ s$^{-1}$ (using 200\,ks pointings)
over an area of $\sim$0.3\,deg$^{2}$.

\cite{ballantyne11} reported the number of expected AGN, in NuSTAR surveys,
as predicted from different population synthesis models. Since
the NuSTAR field of view is smaller than the surveyed area, NuSTAR will 
have to use a tiling strategy. Two tiling strategies can be foreseen:
a corner shift and half-shift survey \citep[see also][]{ballantyne11}.
In the corner-shift strategy, the survey area is covered by non-overlapping
pointing. In the half-shift strategy the distance between pointings
is half the size of the FOV. The corner-shift strategy reaches deeper fluxes
while the half-shift  reaches a more uniform exposure of the surveyed area.
\cite{ballantyne11} concluded that the half-shift strategy yields
the larger number of AGN. Thus, in order to make
a proper comparison we adopt the sky coverages reported by \cite{ballantyne11}
for the half-shift tiling strategy for the 10--30\,keV band.
Converting the sky coverages from the 10--30\,keV to the 15--55\,keV
is rather easy and almost error free. Indeed for the power-law
and the PEXRAV model discussed in $\S$~\ref{sec:counts} we get a conversion
factor of 1.16 and 1.18 respectively. Thus using a LogN-LogS derived
in an overlapping band reduces the uncertainties due to flux conversion
to less than $\sim$2\,\%.

\begin{figure*}[ht!]
  \begin{center}
  \begin{tabular}{cc}
\hspace{-1cm}
    \includegraphics[scale=0.5]{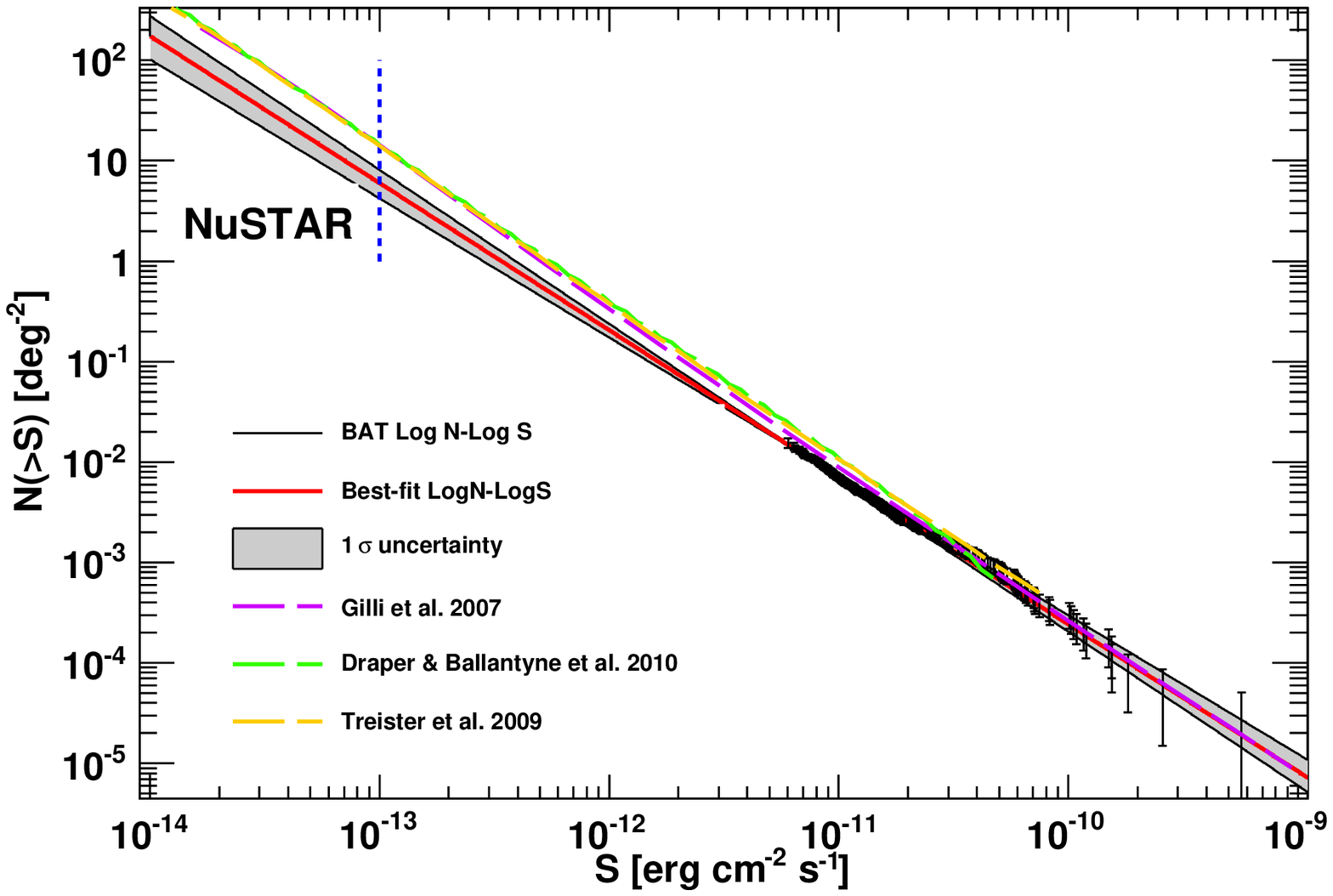}&
\hspace{-1cm}
  	 \includegraphics[scale=0.5]{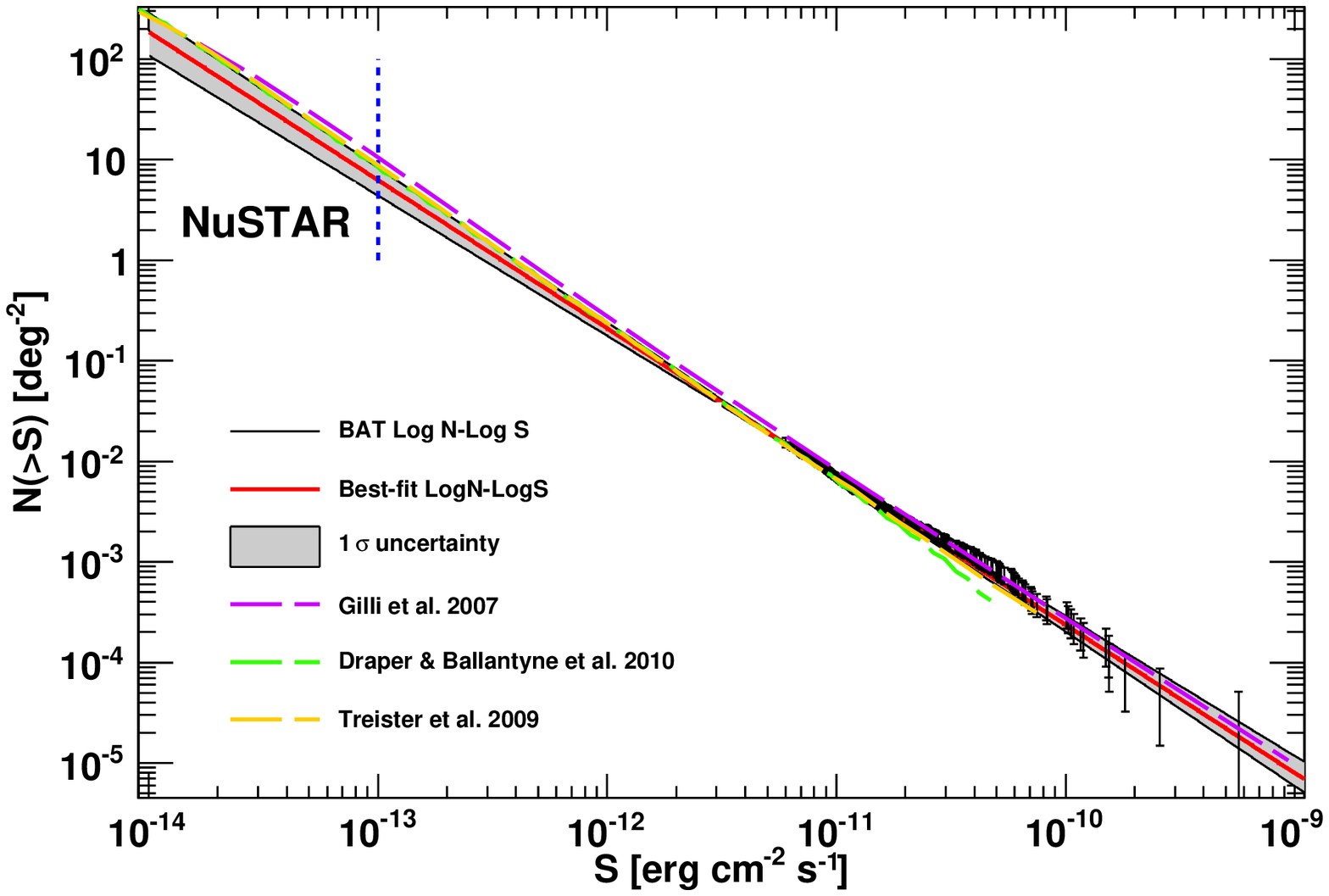}\\
\end{tabular}
  \end{center}
  \caption{LogN-LogS of the BAT AGN (black data points) and best power-law fit
extrapolated to the sensitivity that  NuSTAR will achieve.
The left panel shows the original 
prediction of synthesis models, while in the right panel the predictions
have been renormalized to match the AGN densities measured by BAT.
In both panels the vertical
dashed line shows the sensitivity reached by NuSTAR in a short
pointing (i.e. 5--10\,ks).
\label{fig:nustar}}
\end{figure*}

The number of objects predicted, extrapolating the LogN--LogS,
 for 4 different survey fields are
reported in Tab.~\ref{tab:counts} and they are compared to the predictions
of the models of \cite{treister09} and \cite{draper10} reported 
in \cite{ballantyne11}. It is clear  that our predictions lie substantially
lower than the ones of population synthesis models.
 The left panel of Fig.~\ref{fig:nustar} shows
that (at NuSTAR sensitivities) the predictions of synthesis models lie
a factor $\sim$3 above the 
extrapolation of  the LogN-LogS of BAT AGN.

This disagreement is likely due to two reasons.
Part of it is due to the fact that synthesis models lie systematically
above the BAT LogN--LogS as shown in $\S$~\ref{sec:counts}.
The second one is that the simple extrapolation of the BAT LogN--LogS
is not able to capture the evolution of AGN which is rather well
established \cite[e.g.][]{miyaji01,lafranca05,hasinger05} 
and expected also in the $\geq$10\,keV band.

In order to correct for the first problem (i.e. the over-prediction
of AGN in the BAT band) we arbitrarily renormalize the models
of \cite{gilli07}, \cite{treister09}, and \cite{draper10} to fit
the BAT data at bright fluxes and then we 
convolve them with the same sky coverages
described above. The predictions for the `re-normalized' models
are reported in the lower part of Tab.~\ref{tab:counts} and
are on average compatible within $\sim$1\,$\sigma$ with
the extrapolation of the BAT LogN--LogS (as also visible in the right
panel of Fig.~\ref{fig:nustar}).
This is reflected into a lower number of predicted AGN detections 
for NuSTAR in the  10--30\,keV band.

In order to correct (or to gauge the uncertainty) due to neglecting any
evolutionary effect in the LogN--LogS we rely on the best-fit PLE model
of $\S$~\ref{sec:xlf}. Even if marginally significant we take the 
best-fit parameters at their face value and derive the number of 
expected objects in NuSTAR fields. These are reported in the last
row of Tab.~\ref{tab:counts}. As it can be seen the prediction
from the evolutionary XLF are in fairly good agreement with the 
predictions from the `normalized'  synthesis models.
We thus believe this set of predictions (i.e. lower part of Tab.~\ref{tab:counts})
for the number of AGN detectable is realistic.
However, the ultimate number of detected AGN
will likely depend on how the sensitivity will vary across the survey fields
and, for the smallest fields, on cosmic variance.

\begin{deluxetable}{lcccc}
\tablewidth{0pt}
\tablecaption{Predictions for NuSTAR surveys adopting the half-shift coverages
reported in \cite{ballantyne11}. The lower part of the table shows the prediction
of the models renormalized to match the BAT LogN--LogS.
\label{tab:counts}}
\tablehead{
\colhead{Model}       & \colhead{BO\"{O}TES}                &
\colhead{COSMOS}      &\colhead{ECDFS} & \colhead{GOODS}      \\
%
\colhead{}          & \colhead{\scriptsize (9.3\,deg$^{2}$)}         &
\colhead{\scriptsize (2\,deg$^{2}$)} & \colhead{\scriptsize (0.25\,deg$^{2}$)}
& \colhead{\scriptsize (0.089\,deg$^{2}$)}
}

\startdata

This work (no evo.) & 47$^{+16}_{-13}$ & 31$^{+13}_{-10}$ &20$^{+10}_{-7}$ & 16$^{+8}_{-6}$\\

Draper \& Ballantyne 2010 &  126 & 91 & 62  & 51\\
Treister et al. 2009      &  107 & 77 & 52 & 42\\
\hline
Draper \& Ballantyne 2010 &  66  & 46 & 31  & 26\\
Treister et al. 2009      &  68  & 51 & 33  & 27\\
Gilli et al. 2007         &  94  & 60 & 39  & 25\\
This work (evo.)          &  61  & 42 & 28  & 23\\ 
\enddata 
\end{deluxetable}

\section{Summary and Conclusions}
\label{sec:concl}

The analysis presented here shows the power of an all-sky survey at hard
X-rays to study the AGN in our local environment. 
Thanks to the large field of view and high sensitivity  
BAT has detected 428 AGN all-sky above 10\,keV with negligible ($\leq$5\,\%)
incompleteness. This represents the largest complete sample 
of AGN detected so far.
Below we summarize our findings.
\begin{itemize}
\item The BAT AGN sample spans 10 decades in luminosity comprising objects
detected at distance of  $\sim$1\,Mpc up to redshift $\sim$3.5. The AGN sample
can be divided into objects whose emission is dominate by the accretion disk/corona (i.e. Seyfert galaxies) and jet-dominated sources (i.e. blazars and radio galaxies). Seyfert galaxies are detected at low redshifts and low luminosities 
while jet-dominated sources, that are $\sim$15\,\% of the whole sample,
are detected at high luminosity and high redshift.
\item Samples of AGN detected above 10\,keV are instrumental to determine
the size of the population of Compton-thick AGN that are still detected
in very little numbers. While a detailed measurement of the absorbing column
density of all the AGN in the BAT sample is left to a future publication,
we cross-correlated our sample with known catalogs of bona-fide Compton-thick 
AGN. The BAT sample already comprises 15 \cite[out of the 18 reported
by][]{dellaceca08a} bona-fide Compton-thick AGN and 3 likely candidates.
The observed\footnote{See \cite{burlon11} for the bias that also instruments
above 10\,keV have in detecting Compton-thick AGN.} fraction of Compton-thick
AGN, relative to the whole population, is thus $\sim$5\,\%.
We showed that BAT will likely not detect the rest of the known candidate
Compton-thick AGN. Since the BAT sensitivity still improves with time,
 future AGN samples detected by BAT will likely
contain previously unstudied Compton-thick AGN. A few (1 or 2 objects) might
be present already in this sample.
\item We performed a robust analysis, using bootstrapping,
 of the source counts distribution of the Seyfert-like objects detected by
BAT. The BAT LogN--LogS is consistent with Euclidean down to the lowest
fluxes spanned by this analysis (i.e. 6$\times10^{-12}$\,erg cm$^{-2}$ s$^{-1}$).
The agreement between our and the INTEGRAL results \citep{krivonos10} shows
that the LogN--LogS of AGN selected above 10\,keV is established to a precision of  10\,\%. The population synthesis models that we have
tested \citep[i.e.][]{gilli07,treister09,draper10} are able to reproduce
the BAT LogN--LogS at the brightest fluxes, 
but overestimate it at the lowest fluxes
spanned by this analysis.
\item We derived the luminosity function of (local) AGN and tested
for its possible evolution with redshift. Even with our large sample
of AGN the evidence for the evolution of the XLF are at best marginal 
(i.e. $\sim$2\,$\sigma$). The BAT data are well described by a non-evolving
XLF which is modeled as a standard double power law. We find
that the slope of the faint end is $\gamma_1=$0.79$\pm0.08$  while
that one of the bright end  is $\gamma_2$=2.39$\pm0.12$. These values
are in good agreement with those of \cite{sazonov08} and \cite{tueller08},
but better constrained. Given the small FOV, NuSTAR will not be able to
directly constrain the properties of the low-luminosities low-redshift AGN.
The BAT sample and the BAT XLF are thus instrumental to determine, in connection
with the future NuSTAR samples, the evolution and growth of AGN.
\item We derived the luminosity function of Compton-thick AGN
and found that at redshift zero their space density
is 7.9$^{+4.1}_{-2.9}\times10^{-5}$\,Mpc$^{-3}$
for objects with a de-absorbed luminosity
larger than 2$\times10^{42}$\,erg s$^{-1}$. Our measurement is slightly larger, but compatible within uncertainties, with the prediction of synthesis models.
\item  The BAT samples of Seyfert galaxies is truly a local sample (median
redshift 0.03) and can be used to study the spatial distribution of AGN in the 
local Universe. We detected significant over-density features in the spatial
distribution of AGN located within 85\,Mpc. The densest regions show 
a density of AGN that is up to $\sim$5 times larger than the average all-sky
density. These dense regions can be identified with the most prominent nearby
super-clusters: i.e. the Hydra-Centaurus super-cluster and the `Great Attractor'.
The fraction of Sy2 galaxies (with respect to the total AGN population) 
appear to be  larger
than average in the direction of the over-dense regions. This evidence might
support an evolutionary link where close encounters 
of galaxies trigger AGN activity whose first appearance is obscured by dust and gas.  
\item The BAT and the NuSTAR energy bands overlap and it is thus possible
to derive straightforward predictions, from the LogN--LogS or the XLF,
 for the number of AGN that NuSTAR might detect in survey fields in the near
future. We find substantial agreement in the number of predicted objects
if the predictions of population synthesis models are renormalized to
match, at the lowest fluxes, the BAT LogN--LogS.
\end{itemize}

Owing to the capability of detecting the local (those within $\sim$200\,Mpc), 
relatively low-luminosity AGN, but also high-redshift blazars, the all-sky hard X-ray 
surveys - such as the {\it Swift}-BAT survey discussed in this paper - have 
the unique potential to study simultaneously both the nearby and the early Universe.  
Such hard X-ray surveys represent a unique resource, since for many years, they 
will remain the most sensitive probes of the accretion history in the Universe.

\acknowledgments
MA acknowledges extensive discussions with Nico Cappelluti.
MA acknowledges Ezequiel Treister and David Ballantyne 
for making the predictions from their models
available in a machine readable format. The authors acknowledge
the comments of the referee which helped improving this paper.

{\it Facilities:} \facility{Swift/BAT}

\bibliographystyle{apj}

\end{document}